# Carbon Decorated $TiO_2$ Nanotube Membranes: A Renewable Nanofilter for Size- and Charge Selective Enrichment of Proteins

Jingwen Xu, Lingling Yang, Yuyao Han, Yongmei Wang, Xuemei Zhou, Zhida Gao, Yan-Yan Song,* Patrik Schmuki*

**Abstract**

In this work, we design a $TiO_2$ nanomembrane (TiNM) that can be used as a nanofilter platform for a selective enrichment of specific proteins. After use the photocatalytic properties of $TiO_2$ allow to decompose unwanted remnant on the substrate and thus make the platform reusable. To construct this platform we fabricate a free-standing $TiO_2$ nanotube array and remove the bottom oxide to form a both-end open TiNM. By pyrolysis of the natural tube wall contamination (C/TiNM), the walls become decorated with graphitic carbon patches. Owing to the large surface area, the amphiphilic nature and the charge adjustable character, this C/TiNM can be used to extract and enrich hydrophobic and charged biomolecules from solutions. Using human serum albumin (HSA) as a model protein as well as protein mixtures, we show that the composite membrane exhibits a highly enhanced loading capacity and protein selectivity and is reusable after a short UV treatment.

## 1. Introduction

The separation of specific protein species from complex biological matrices is one of the crucial challenges in proteomics.[1] In this context, nano-sized sorbents have attracted a lot of attention in solid-phase extraction (SPE) technology.[2] The ideal SPE sorbent should not only have an excellent biocompatibility, a high adsorption capacity and a selective protein recognition, but it would also be of a considerable advantage for device design if the platform can be reused after elution of the loaded protein.[3] In recent years, considerable attention has focused on graphitic carbon materials, i.e. mesoporous carbon,[4,5] graphene,[6] and carbon nanotubes,[7,8] owning to their delocalized $\pi$-electron system and large surface area that can provide strong affinity sites for a wide range of proteins.[9] In practice, however, the application of graphitic carbon materials alone is cumbersome due to the relative difficulty in isolating and recovering the nano-sized sorbent from the mixed suspension, which frequently leads to secondary contamination.[10] To tackle this problem, carbon materials are usually attached on various scaffolds using silane linkers to form composite nanostructures.[11-14] For example, graphene decorated silica has been used for specific adsorption to phosphorylated peptides,[6] or graphene-bound $Fe_3O_4$ nanoparticles have been applied in bovine serum albumin (BSA) adsorption.[15]

Over the past decade a number of ‘‘oxide’’ nanotubular or nanoporous systems grown by a self-organizing electrochemical anodization have attracted tremendous scientific and technological interest.[16] Such self-organized regular arrays are grown aligned and perpendicular to the substrate surface. This arrangement makes such structures ideal for loading, capturing, and concentrating of secondary species. Most frequent structures are based on silica, titania and alumina.[17-19] Among these materials, $TiO_2$ nanaotube arrays are most attractive for reusable devices due to the unique photocatalytic properties of $TiO_2$, which make these arrays have self-cleaning features.[20] Additionally, nanotubes of $TiO_2$ provide excellent biocompatibility.[21] However, for protein adsorption, $TiO_2$ nanotube walls should be modified preferably with carbonaceous layers. Nevertheless it is difficult to introduce carbon materials (i.e. carbon nanotubes, graphene and carbon dots) into the nanotube without using any silane or phosphonic acid linker, [22] and at the same time keep the tube mouth still open. Moreover, the linkers normally used in preparation of the sorbent are unfavorable for “green” solid-phase extraction. Until now, no investigation has directly applied suitable carbon modified $TiO_2$ nanotubes as sorbents for selective protein separation or enrichment from biological matrices.

In this study, we prepare a free-standing $TiO_2$ nanomembrane (TiNM) by removing the bottoms of aligned $TiO_2$ nanotube arrays. A facile pyrolysis approach is applied to graft a layer of carbon on nanotube wall. The formed hybrid membrane (C/TiNM) combines the characters of $TiO_2$ and carbon, and is then applied as nanofilter in selective recognition and enrichment protein from protein mixtures. The practical applicability of the membrane is validated by selective extraction of HSA from human blood. Due to the unique photocatalytic properties of $TiO_2$ these membranes additionally exhibit excellent self-cleaning features, which makes them attractive for reusable devices.

For the assembly of the platform, in a first step self-organized $TiO_2$ nanotube arrays were grown by anodization of Ti foils in ethylene glycol electrolytes containing $H_2O_2$ and $NH_4F$ as described in the experimental section (supporting information). Fig. 1a-c show SEM images of the as-formed $TiO_2$ nanotube arrays. Under the present conditions a ~55 μm-thick layer of $TiO_2$ nanotubes with openings of 180±15 nm was obtained. These layers can be lifted-off from the metallic substrate and a free-standing nanotube membrane (TiNM) as shown in Fig. 1b is obtained. In order to open the closed bottoms of the members the bottom part was exposed to HF fumes (as described in more details in the experimental section). Fig. 1c and Fig. 1d show the bottom of TiNM before and after HF treatment. From image analysis it can be estimated that the treatment leads to ~90% tube bottoms that are opened (Fig. S1†). To demonstrate the successful opening of the channels we placed a droplet of RhB solution onto the top of the free-standing TiNM and observed permeability. Fig. S2 shows that the red solution penetrated the membrane after several minutes, i.e. open channels are present (this coloration of the bottom does not occur if no HF-fume treatment is applied).

As the nanotube arrays are formed by electrochemical anodization in an organic electrolyte − ethylene glycol (EG) − the tube walls, if not rinsed remain decorated with a thin EG film. This can be used advantageously to coat a carbonaceous (graphite like) film (Fig. 1e and 1f) on the nanotube walls. This conversion of the EG layer by a simple pyrolysis process in inert gas[23] results in changes in the morphology (Fig. 1e and 1f, Fig. S3†). The formed carbon residues are present as regular patches on the tube wall from top to bottom (Fig. 1g-i). Since the carbon layer is hydrophobic, while $TiO_2$ is hydrophilic, the resulting C/TiNM based nanocomposite exhibits amphiphilic properties. XRD, Raman and XPS measurements further confirm the presence and nature of the carbon layer. In Fig. 2a, the diffraction peaks of anatase $TiO_2$ and carbon can be clearly observed in the XRD pattern of the C/TiNM sample. The Raman spectra in Fig. 2b present two new peaks at 1346 $cm^{-1}$ and 1586 $cm^{-1}$ for C/TiNM, which can be attributed to the defect-induced band (D band) and ordered graphitic structure (G band) of the carbon layer, respectively.[24] Fig. 2c exhibits the XPS spectra of C 1s before and after heat treatment in $N_2$ (a comparison to neat NTs and carbon decorated NTs is shown in Fig S4†). Before the pyrolysis process, the C 1s signal can be fitted with four peaks at 284.4 eV, 285.1 eV, 287.6 eV and 288.2 eV, which correspond to C=C, C–O, C=O and -COOH bonds respectively. The apparent increase of C=C peak and decrease of C–O peak after annealing in $N_2$ indicates the formation of partly graphic carbon by decomposing of residual EG molecules. Moreover these peaks corresponding to C 1s vanish after annealed in air as carbon is oxidized and burnt off in $O_2$ (Fig. S4†). In addition, the decrease of the O 1s peaks (Fig. 2d) at 530.7 eV (C=O) and 532.7 eV (C-O-C) further demonstrate the decomposition of organic molecules by the heating treatment.

These layers (C/TiNM) were explored as sorbent for the selective adsorption, enrichment and separation of proteins. To evaluate the selectivity provided by C/TiNM, experiments with human serum albumin (HSA, pI 4.9), cytochrome c (Cyt-c, pI 9.8) and hemoglobin (Hb, pI 6.8−7.0) were carried out. For absorption measurements the membranes and protein solutions were mixed for some time and the remaining protein in the solution was determined. For HSA, the absorbance at 595 nm of the Coomassie Brilliant Blue dye after binding with HSA (Fig. S7a and S7b) was measured, as described in detail in the experimental section. As shown in Fig. 3a, the adsorption efficiencies of the C/TiNM towards the different proteins are pH dependent. The favorable pH values for protein adsorption are close to the isoelectric points (pI) of each protein, i.e., pH 5.0 for HSA, pH 7.0 for Hb and pH 10.0 for Cyt-c. To demonstrate selective enrichment we prepared a mixture of HSA, Hb, Cyt-c and C/TiNM and carried out SDS-PAGE assays as described in experimental section. The results in the Fig. 3b show that HSA can be separated from the mixture of HSA, Hb and Cyt-c successfully using adsorption at pH 5.5 and elution at pH 8.8. In analogy, Hb and Cyt-c can also be separated from mixtures sing corresponding proper pH values (Fig. S5†).

To better understand the protein adsorption, C/TiNM, HSA was selected for further studying the enrichment mechanism and extracting characteristics in more detail. Binding of HSA on C/TiNM can be observed in FT-IR spectra (Fig. S6†) – this is most evident from two new bands that appear at around 1628 $cm^{-1}$ and 1572 $cm^{-1}$ of the C/TiNM after the adsorption of HSA molecules.

The ionic strength is one of the important factors that influence protein adsorption. In Fig. S7c†, the adsorption efficiency is noted to increase slightly with the addition of NaCl and achieve the maximum value at a NaCl concentration of 0.05 mol $L^{-1}$. The salt molecules can coordinate with the water shell surround the HSA, and thus facilitate the exposure of hydrophobic residues in protein molecules.[25] However, at high salt concentration, salt molecules compete with the protein for the available adsorption sites which leads to a decrease of the adsorption ability. In the following we used a NaCl concentration of 0.05 mol $L^{-1}$ to control the ionic strength of the sample solution. The adsorption kinetics curve of C/TiNM shows that the highest adsorption efficiency is reached when the adsorption time is ~180 min (Fig. S7d†). The adsorption isotherms of HSA on C/TiNM and TiNM were measured by performing the adsorption process from a B-R buffer (pH 5.0) containing 0.05 mol $L^{-1}$ NaCl at 25 °C within a range of initial protein concentration (20−160 mg $L^{-1}$). In Fig. 3c, noticeably, the C/TiNM samples present a significantly improved adsorption ability (the saturated adsorption capacity of HSA is 41.8 mg $g^{-1}$) compared to the bare TiNM sample (TiNM shows a saturated adsorption capacity of ~16.7 mg $g^{-1}$) under the same experimental conditions. However, the BET measurements only show a minor increased surface area from TiNM (Fig. S8†, 23.7 $m^2$ $g^{-1}$) to C/TiNM (Fig. S8†, 25.6 $m^2$ $g^{-1}$). Thus the improvement in protein extraction can be mainly attributed to the difference in the tube’s surface character. As plotted in Fig. 3d, the ζ potentials of C/TiNM and TiNM are recorded in a series of B-R buffer with different pH values. Obviously, the C/TiNM samples present a positively charged surface in acidic solutions (pH<3.5). When the solution pH is lower than the isoelectric point of HSA (pI 4.9), HSA molecules are also positively charged. In this case, the strong electrostatic repulsion between protein and C/TiNM leads to the poor absorption efficiency at these low pH values. When the solution pH is close to the isoelectric point (pI) of HSA, the C/TiNM exhibits the highest adsorption efficiency. At this point, protein molecules are uncharged, thus the electrostatic interaction with the tube wall becomes weak. At the same time, more embedded hydrophobic residues in the proteins are exposed, and the carbon layer is hydrophobic, which can result in strong hydrophobic interactions between the protein molecules and the carbon patches and therefore this effect provides an alternative driving force for protein adsorption.[9] As shown in Fig. 3a, with a further increase of the pH (pH>pI), the adsorption efficiency decreases due to the enhanced electrostatic repulsion, because the protein and C/TiNM are both negatively charged (Fig. 3d).

The feasibility to change the adsorption behavior of C/TiNM by a variation of the pH over a wide range provides a high potential to a use for the selective extraction of a considerable number of target proteins from mixtures of proteins or complex sample matrices. On the other hand, the bare TiNM samples process a negative $\zeta$ potentials over a wide pH range (from pH 2 to pH 10). Without carbon layer grafting, the electrostatic interaction is the main driving force for protein adsorption and thus when the pH value is close to the pI value of HSA, the electrostatic interaction becomes week. Above results thus demonstrate the advantages of using C/TiNM for isolating target proteins – as this simply can be done by adjusting the pH value of a buffer solution.

Except for adsorption also elution of the loaded protein from the adsorbent, by an appropriate stripping agent, is important for a use in biological applications. In this respect, several common used buffer solutions were considered as the stripping agents to elute HSA from C/TiNM. As evident from Fig. 4a, a favorable recovery is achieved (with elution efficiencies of ~90%) when alkaline buffers are employed as a stripping reagent. According to the ζ potentials of C/TiNM (Fig. 3d), buffer solutions with a higher pH value can generate electrostatic repulsion between C/TiNM and HSA (both are positively charged), which leads to the detachment of HSA from C/TiNM. By further considering pH ranges that avoids unfavorable effects on the protein conformation and biological activity, it is preferential to use B-R or PBS buffer as a stripping reagent. The comparison in Fig. S9a† demonstrates the B-R buffer has a better performance for HSA elution over a wide pH range.

In further steps the effect of ionic strength and pH value on the recovery efficiency was considered. The protein recovery efficiency declines with the addition of NaCl (Fig. S9b†). The B-R buffer solution of pH 8.8 shows the best performance (Fig. S9c†) and a recovery efficiency of ~92% is obtained under optimized conditions (Fig. S9d†). To evaluate if denaturation effects can be induced by the adsorbent, circular dichroism (CD) spectra were employed to analyze the HSA recovered from C/TiNM. In Fig. 4b, the two negative absorption bonds in far-UV region at around 220 and 209 nm are assigned to the typical feature of n–π* transfer of the peptide bond in the α+β-helix of HSA.[26] A comparison of the overlap of the CD spectra of the recovered HSA and standard HSA indicates that the secondary structure of HSA remains intact during the adsorption/desorption processes – this further suggests that C/TiNM is well suitable for a use as non-denaturalizing adsorbent for protein enrichment.

Reusability is another factor for an adsorbent in practical applications. To evaluate photocatalytic cleaning and repeated adsorption on C/TiNM samples, we carried out recycling tests. Each sample was reused in five successive adsorption–desorption cycles. In one set of samples after the elution of the protein, the C/TiNM was then cleaned by exposure to UV light for some minutes in another set this photocatalytic cleaning left out. As shown in Fig. 4c, after five cycles, the samples without UV-cleaning keeps only ~20% of its initial absorption efficiency while the samples illuminated in each cycle keeps approx. ~85% of its initial absorption efficiency.

We additionally confirmed the photocatalytic activity of the C/TiNM using a standard photocatalytic test. Fig. S10† shows the photocatalytic degradation of RhB on C/TiNM and TiNM. This shows that the carbon decorated TiNM to provide similar photocatalytic activities as bare TiNM. As shown in XRD results (Fig. 2a), C/TiNM nanocomposite consists of anatase $TiO_2$ which is well-known for its photocatalytic self-cleaning activities.[27] From TEM images (Fig. 1), discrete amorphous carbon layer can be observed. The lattice fringe of $TiO_2$ nanocrystal (Fig. 1 g-i) is determined to be 0.351 nm and 0.190 nm, which corresponds to the anatase phase (101) and (200) respectively. I.e. this patchy carbon coverage of the C/TiNM wall leaves bare $TiO_2$ partly exposed and thus active for photocatalytic reactions, that is the carbonized surface aids anchoring the proteins and the bare surface provides photocatalytic activity. The cycling experiments show that organic molecules (i.e. pollution, biomolecules) in the tubes can be decomposed by UV illumination, but adsorption sites (graphitic carbon) remain active for a next protein absorption.[28] Furthermore, we also recorded the mass of samples during the absorption-elution cycles. As shown in Fig. 4d, the three C/TiNM samples don't exhibit obvious lost in weight for repeated adsorption-elution if UV-cleaning treatments are used. These results suggest the C/TiNM nanocomposite has excellent recyclability in adsorption–desorption cycles and does not loose significant amounts of the decorated carbon in the cleaning cycles.

To illustrate that C/TiNMs can be used as a *practical* nanofilter for protein extraction, we carried out further tests. First, the as-prepared membrane was arranged in a holder and placed in front of a syringe (inset, Fig. 5a). After the HSA solution (100 mg $L^{-1}$) has passed through the nanofilter, the solution concentration shows a significant decrease to 20 mg $L^{-1}$. (Fig. 5a). Second, we used the C/TiNM nanocomposite for the isolation of HSA from human whole blood samples. The human whole blood was donated by a healthy volunteer from the Northeastern University Hospital, Shenyang. Fig. 5b shows the SDS-PAGE assay conducted according to the method reported by Laemmli.[29] On lane II, a few protein bands are observed for the human whole blood sample, which are attributed mainly to human serum albumin (HSA), immunoglobulin G (Ig G), Cyt-c, transferrin (Trf) and Hemoglobin (Hb). After isolation treatment by C/TiNM (details are given in the experimental section), the HSA band (at ca.66.4 kDa) becomes much lighter while the other band still remains at the same intensity in Lane III. After the stripping process by B-R buffer (pH 8.8), a single band of HSA is clearly observed on Lane V. These results indicate that HSA can be selectively isolated by C/TiNM even from a complex real sample, where a large number of factors and proteins could potentially interfere.

In conclusion, we developed a binary nanomembrane (C/TiNM) which exhibits excellent selective enrichment properties toward proteins. The patchy carbon layer within a $TiO_2$ nanotube array is suggested to provide a hydrophobic interface to interact with hydrophobic parts of proteins, and thus facilitates the selective adsorption and desorption of proteins. Such C/TiNM can thus be employed as a nanofilter for enrichment and selective isolation protein from complex matrices. Moreover C/TiNM samples can be easily refreshed and reused due to the photocatalytic activity of $TiO_2$. We believe that the principle shown here, i.e. to use a patchy graphitized nanomembrane has the potential to be used in a much wider range of capturing and release schemes in self-cleaning nanoreactor applications.

**Experimental Section**

*Materials*: Hemoglobin (H2625), human serum albumin (A1653) and cytochrome (30398) were purchased from Sigma-Aldrich (St. Louis, USA) and used as received. Ti foils (0.1 mm thickness, 99.6% purity) was purchased from Advent Research. $NH_4F$, glycerol, $H_2O_2$, sodium dodecyl sulfate (SDS), Coomassie brilliant blue (CBB) and other chemicals were purchased from Sinopharm Chemical Reagent Co. Ltd. and used without further purification. The protein molecular weight marker (broad, 3452, Takara Biotechnology Company, Dalian, China) is a mixture of nine purified proteins ($M_r$ in kDa: myosin, 200; $\beta$-galactosidase, 116; phosphorylase B, 97.2; serum albumin, 66.4; ovalbumin, 44.3; carbonic anhydrase, 29; trypsin inhibitor, 20.1; lysozyme, 14.3; aprotinin, 6.5).

Britton-Robinson (B-R) buffers were employed for the investigation of the protein adsorption behavior. Typically, 0.04 mol $L^{-1}$ acetic acid, phosphoric acid, and boric acid are mixed together and the pH values are adjusted by a 0.2 mol $L^{-1}$ NaOH solution. Double distilled (DI) water (>18 MΩ) was used to prepare aqueous solutions.

*Preparation of C/TiNMs*: Ti foils were cleaned in ethanol and DI water by sonication, and then dried with a nitrogen stream. The cleaned Ti foils were pressed together with a Cu plate against an O-ring sealed opening in an electrochemical cell (with 0.5 $cm^2$ exposed to the electrolyte) and then anodized in a ethylene glycol electrolyte containing 0.5% $NH_4F$ and 3% $H_2O_2$ at 80 V for 10 h at room temperature. Ti foils served as the working electrode, and a platinum sheet was used as the counter electrode. After anodization, TiNTs were rinsed by ethanol and then an ultrasonic treatment was employed to remove the "nanograss" on the surface. Subsequently, the samples were immersed in $H_2O_2$ (30 %) until the $TiO_2$ nanotube arrays lifted from the Ti substrate. The resulting $TiO_2$ membrane was then kept in a HF vapor for 20 min to etch the tube bottom, and carefully washed with DI water for 5 min. To prepare carbon decorated TiNM (C/TiNM), the bottom etched nanomembranes were annealed in $N_2$ at 450 °C for 2 hours. To prepared bare TiNM samples, the samples were dipped into ethanol/$H_2O$ (1:1) solution for 36 h, and then dried in oven at 80 $^oC$ for 2 h.

*Adsorption and elution experiments*:

In general, 3.0 mg of the C/TiNM sample was mixed with 1.0 mL of protein sample solution and the mixture was shaken vigorously to facilitate the protein adsorption. To quantify the remaining protein, the supernatant of the suspension was collected after centrifugation at 8 000 rpm for 5 min. The proteins left in the upernatant solution were measured by monitoring the absorbance at 406 nm for Hb, 408 nm for Cyt-c and 595 nm for HSA (afterbind with the Coomassie Brilliant Blue dye according to Bradford method).[30]

After the adsorption process, the C/TiNM was pre-washed with 3 mL of the DI water to remove the loosely adsorbed molecules on the adsorbent surface. Afterwards, 1 mL of stripping solution was added and the solution then slightly shaken for 140 min to facilitate the elution of the adsorbed proteins from the nanotubes. The supernatant containing the recovered protein species was finally collected by centrifugation for the ensuing investigations..

The adsorption efficiency ($E_a$) and elution efficiency ($E_e$) are calculated according to the following equations.

$$E_a = \frac{C_0 - C_1}{C_0} \qquad (1)$$

$$E_e = \frac{C_2}{C_0 - C_1} \qquad (2)$$

$C_0$ is the concentration of HSA in original solution. $C_1$ is the concentration of HSA remained in supernatant after adsorption. $C_2$ is the concentration of recovered HSA in supernatant after eluting process.

The adsorption capacity of C/TiNM or TiNM is calculated according to the equation 3:

$$Q = \frac{(C_0 - C_1) * V}{m} \quad (3)$$

$C_0$ is the initial HSA concentration (mg $L^{-1}$), $C_1$ is the HSA concentration remained in supernatant after adsorption (mg $L^{-1}$), $V$ is the volume of HSA solution (mL), and $m$ is the weight of sorbent (mg).

*Self-cleaning of membrane*: Briefly, 3 mg absorbent was added to 1 mL HSA solution (140 mg $L^{-1}$, 0.01 mol L B-R buffer, pH 5.0) at room temperature. After the adsorption process, the absorbent was centrifuged and elute from the C/TiNM surface by 1 mL of B-R buffer (0.01 mol $L^{-1}$). To conduct the UV-induced self-cleaning process, the absorbent was illuminated by an Hg lamp (300 W) for 60 min, and followed by washing with DI water for three times and drying in an oven at 70 $^{o}$C.

*Selective adsorption of specific protein from protein mixture*: The feasibility of the separation of specific protein was demonstrated by isolating individual protein from a protein mixture. Briefly, a solution of Hb, Cyt-c and HSA (200 mg $L^{-1}$ respectively) was mixed at different pH. For example, to selectively absorb HSA, Hb, cyt-c, the pH of solution was controlled at 5.5, 6.9, and 9.8, respectively. Then 3.0 mg of the C/TiNM sample was added into 1.0 mL of the protein mixture solution to extract the target protein. The obtained target protein was then inactivated by boiling to carry out the SDS-PAGE assay.

*SDS-PAGE experiment*: The SDS-PAGE assay was carried out by using 12% polyacrylamide resolving gels, 5% polyacrylamide stacking gels, and standard discontinuous buffer systems according to Laemmli method.[29] Under a constant voltage at 80 V for hours, bands of different proteins were separated, and the protein bands were then visualized by staining with 0.2% Coomassie Blue G250 for 2 hours.

*Selective adsorption of HSA from human whole blood*: Human whole blood was obtained from the Hospital of Northeastern University. Before adsorption experiment, the whole blood was diluted 100-fold by B-R buffer solution (pH 5.0). After the adsorption process, the surface of C/TiNM was cleaned by DI water for three times. The B-R buffer solution at pH 8.8 was chosen as striping solution. The mixture was shaken vigorously for 3 h and then centrifuged to separate the target protein. The collected proteins were further analyzed by sodium dodecyl sulfate polyacrylamide gel electrophoresis (SDS-PAGE) which was performed by using 12% resolving gel and 5% stacking gel with standard discontinuous buffer systems.

*Characterization*: The morphology of membranes were characterized using a field-emission scanning electron microscope (Hitachi FE-SEM S4800, Japan) and transmission electron microscopy (TEM, JEOL 2000). XRD patterns were acquired on an X'Pert X-ray diffraction spectrometer (Philips, USA). Fourier transform infrared (FT-IR) spectra were recorded on a Nicolet-6700 FT-IR spectrophotometer (Thermo, USA) and the samples were prepared by the KBr pellet method. The UV-vis absorption spectra were measured on a spectrophotometer (Perkin–Elmer, Lambda $XLS^+$, USA). The $\zeta$ potentials were analyzed by Zetasizer Nano ZS90 (Malvern, England). X-ray photoelectron spectra (XPS) were recorded on a Perkin–Elmer Physical Electronics 5600 spectrometer.

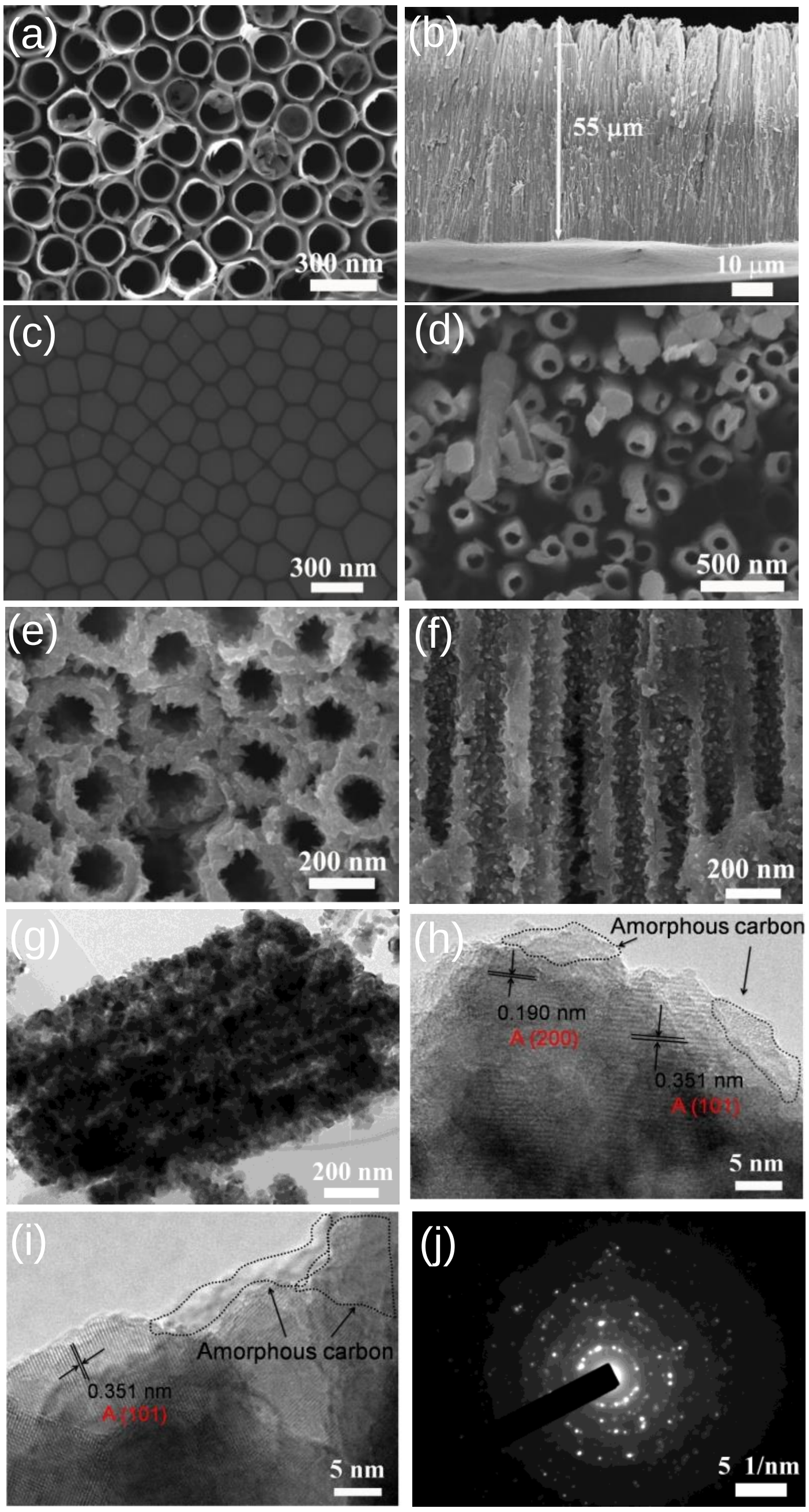

**Fig. 1** SEM images of bare TiNM a) top view, b) side view (inset: the free-standing membrane), bottom view of nanotubes c) before and d) after HF etching; e) top view and f) cross-sectional view of the carbon decorated nanotubes prepared by annealing in $N_2$. g) TEM image of C/TiNM; h) and i) HR-TEM image of C/TiNM representing the nanotubes are composed by anatase crystalline of and discrete carbon layer; j) electron diffraction pattern of C/TiNM.

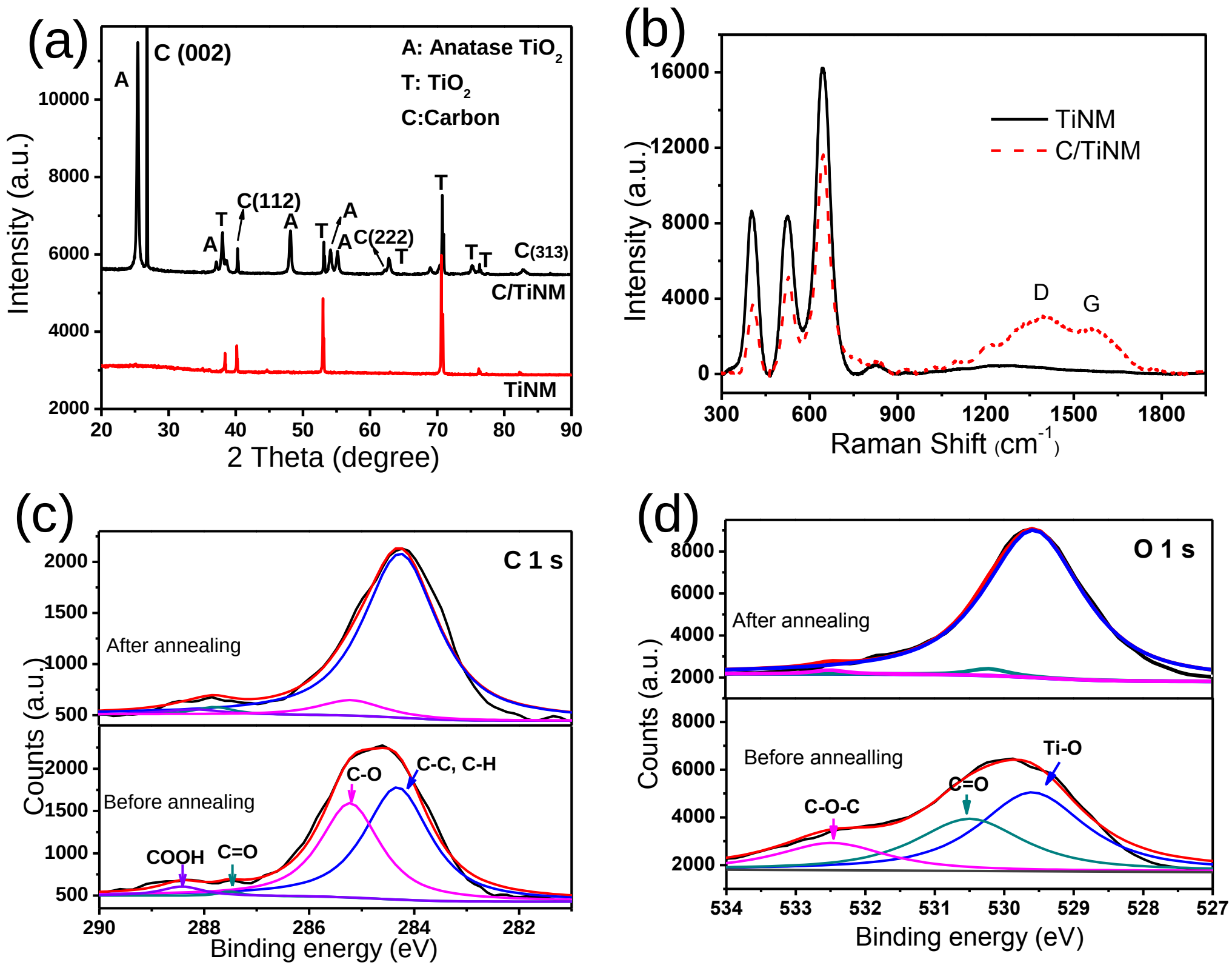

**Fig. 2** a) XRD patterns and b) Raman spectra of TiNM and C/TiNM. XPS high resolution of c) C 1s peak and d) O 1s peak of C/TiNM before and after annealed in $N_2$.

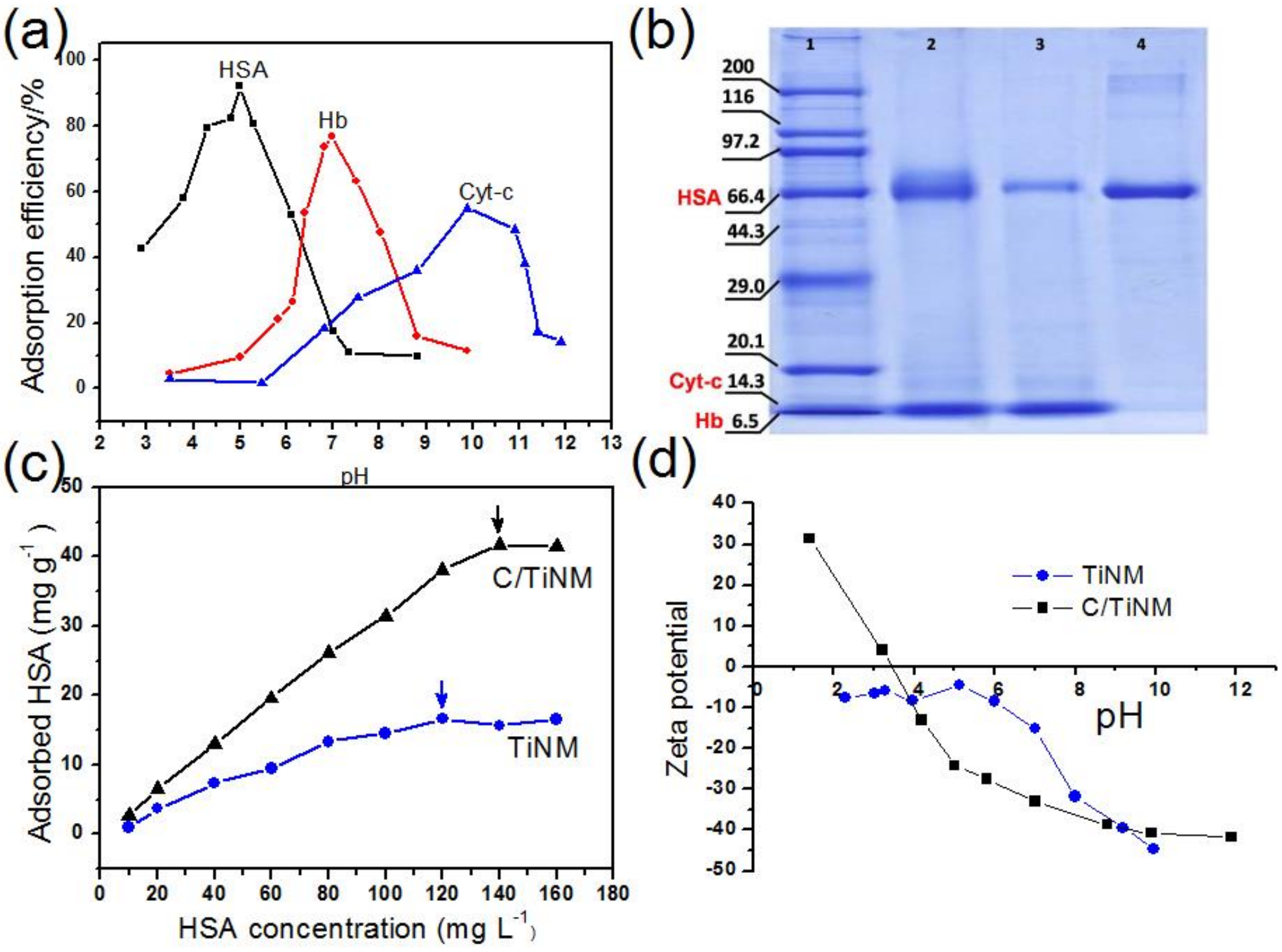

**Fig. 3** a) The adsorption efficiency of HSA, Hb and Cyt-c on C/TiNM at different pH values. b) SDS-PAGE[30] assay results. Lane 1, molecular weight standards (Marker in kDa); lane 2, the mixture of Cyt-c, Hb, and BSA solutions of 400 mg $L^{-1}$; lane 3, the mixture solutions after HSA isolated by C/TiNM (pH=5.5); lane 4, the recovered HSA from C/TiNM (pH=8.8). c) Saturated adsorption isotherms for HSA adsorbed on C/TiNM and bare TiNM. d) The effect of pH on the ζ potential of C/TiNM and bare TiNM.

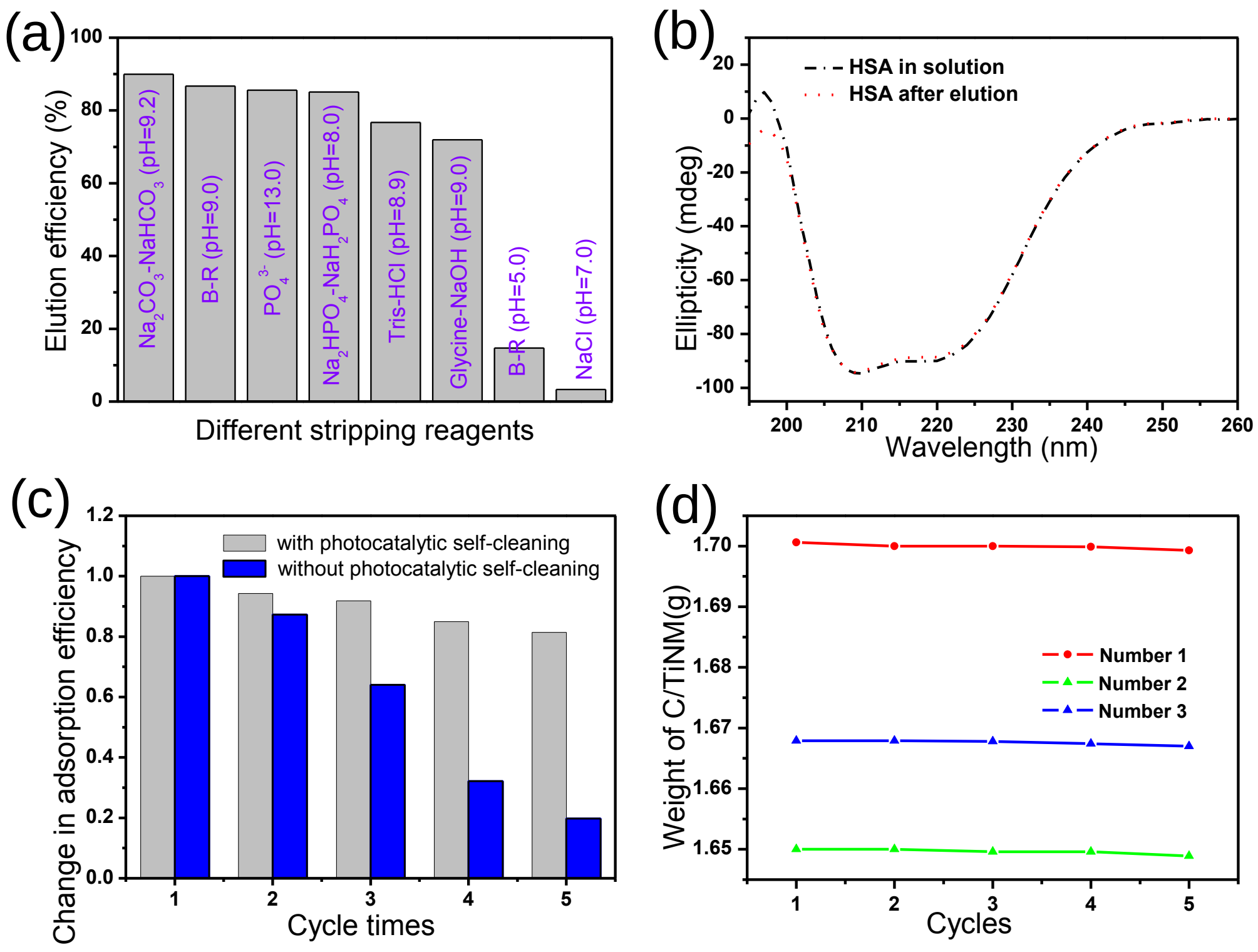

**Fig. 4** a) Elution efficiency of HSA from C/TiNM by using different eluting agents. b) CD spectra of HSA standard solution (100 mg $L^{-1}$) in 0.01mol/L B-R buffer at pH 8.8 and the recovered HSA through adsorption-desorption cycle. c) The change of adsorption efficiency of C/TiNM with and without undergoing UV-irradiation treatment. d) The weight change of C/TiNM samples in five adsorption-desorption cycles.

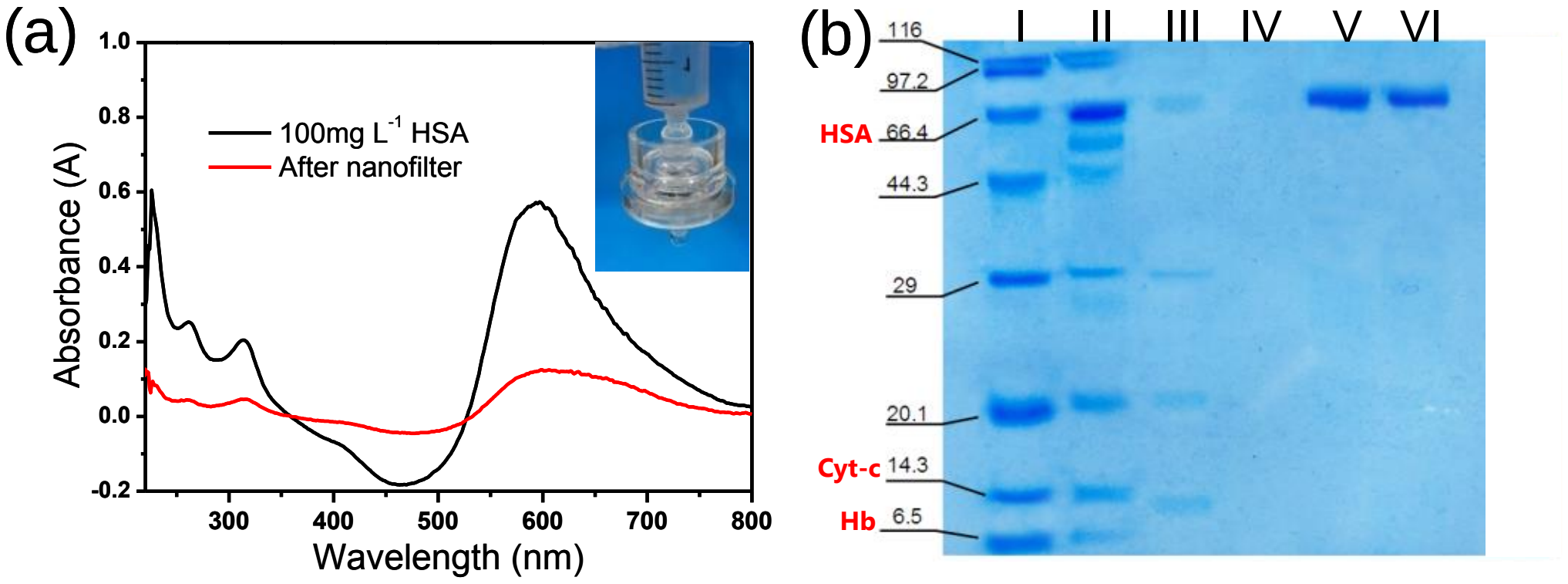

**Fig. 5** a) UV-Vis spectra of 100 mg $L^{-1}$ HSA before and after pass through C/TiNM. b) SDS-PAGE assay results. Lane I, molecular weight standards (Marker in kDa); lane II, 100-fold diluted human whole blood without pretreatment; lane III, 100-fold diluted human whole blood after adsorption by C-TiNM; lane IV, washing solution by deionized water; lane V, HSA recovered from the C-TiNM composite; Lane VI, HSA standard solution of 140 mg $L^{-1}$.

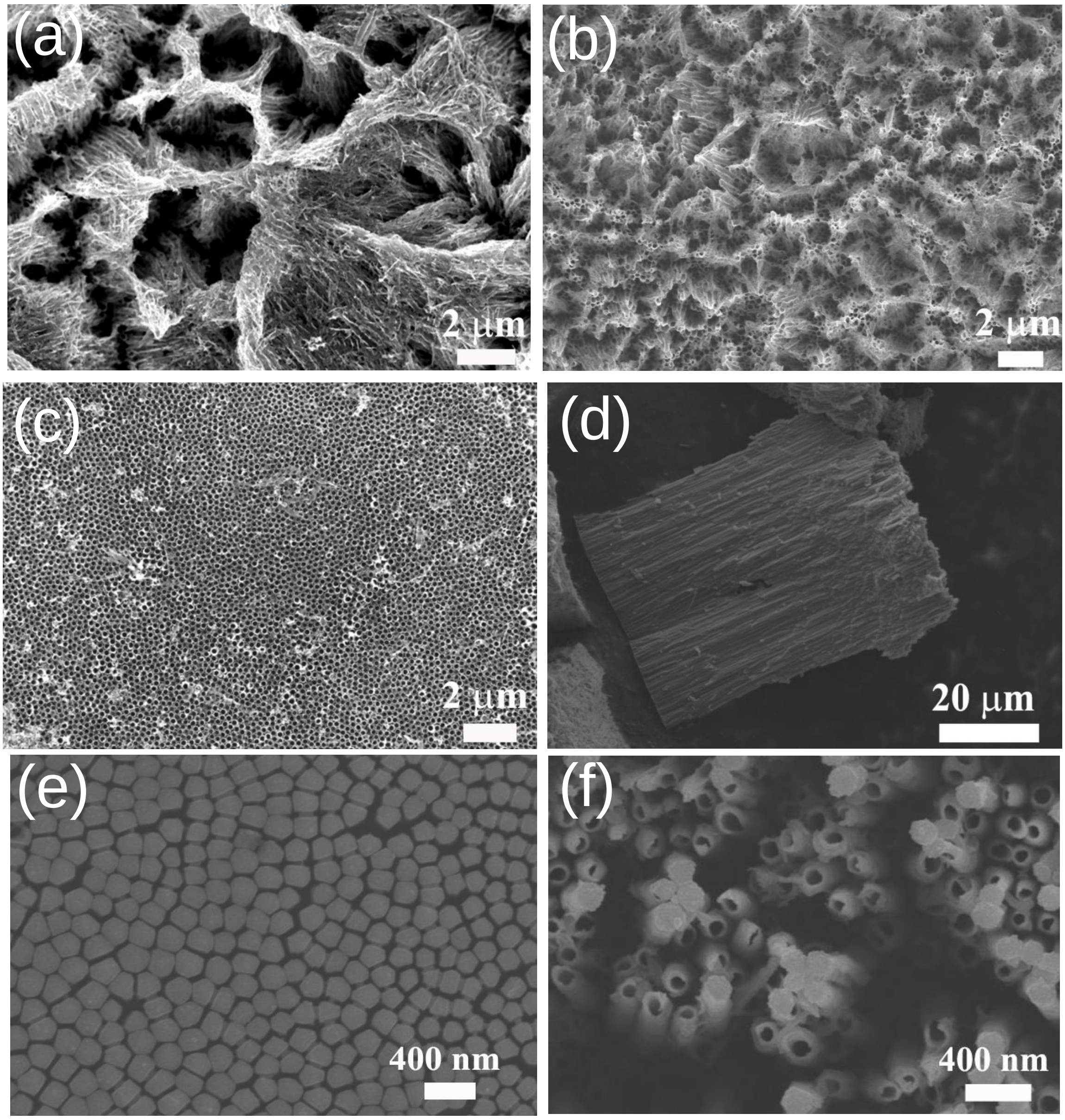

**Figure S1.** SEM images: a) top view of as-prepared bare TiNM, after ultrasonic treatment in ethanol for b) 5 min and c) 15 min in ethanol; d) side view of the bare TiNM; and bottom view of nanotube arrays after etched in HF atmosphere for e) 5 min and f) 10 min.

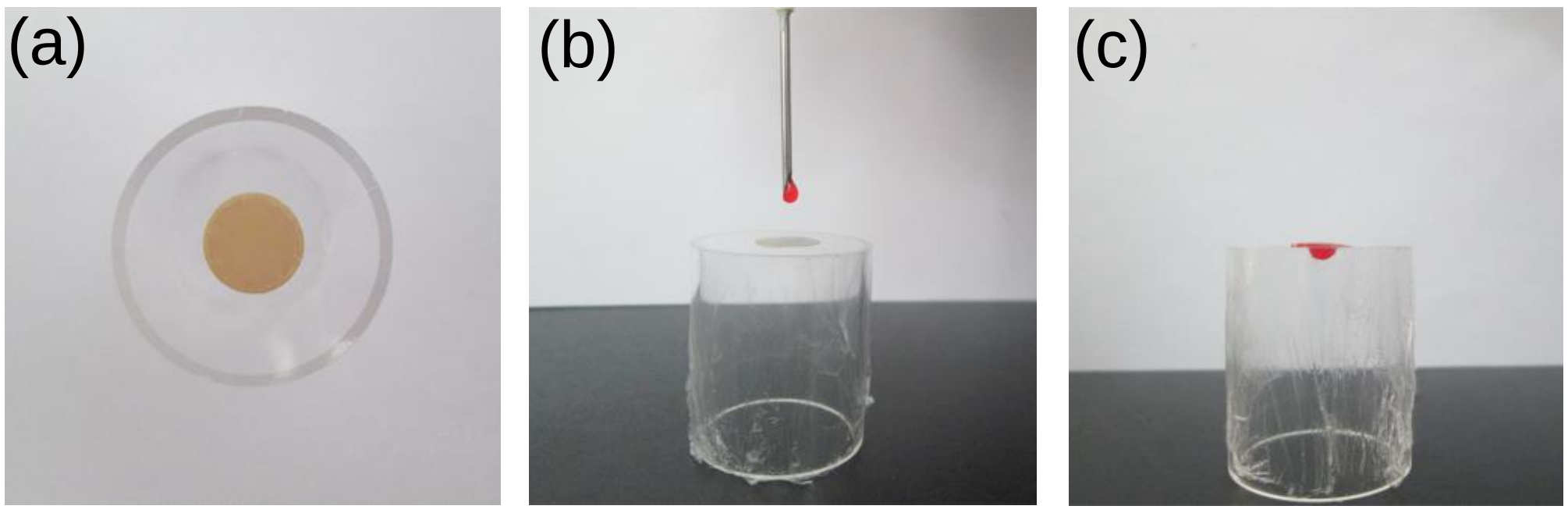

**Figure S2.** The optic diagrams for dye solution (Rhodamine 6G) passing through the TiNM.

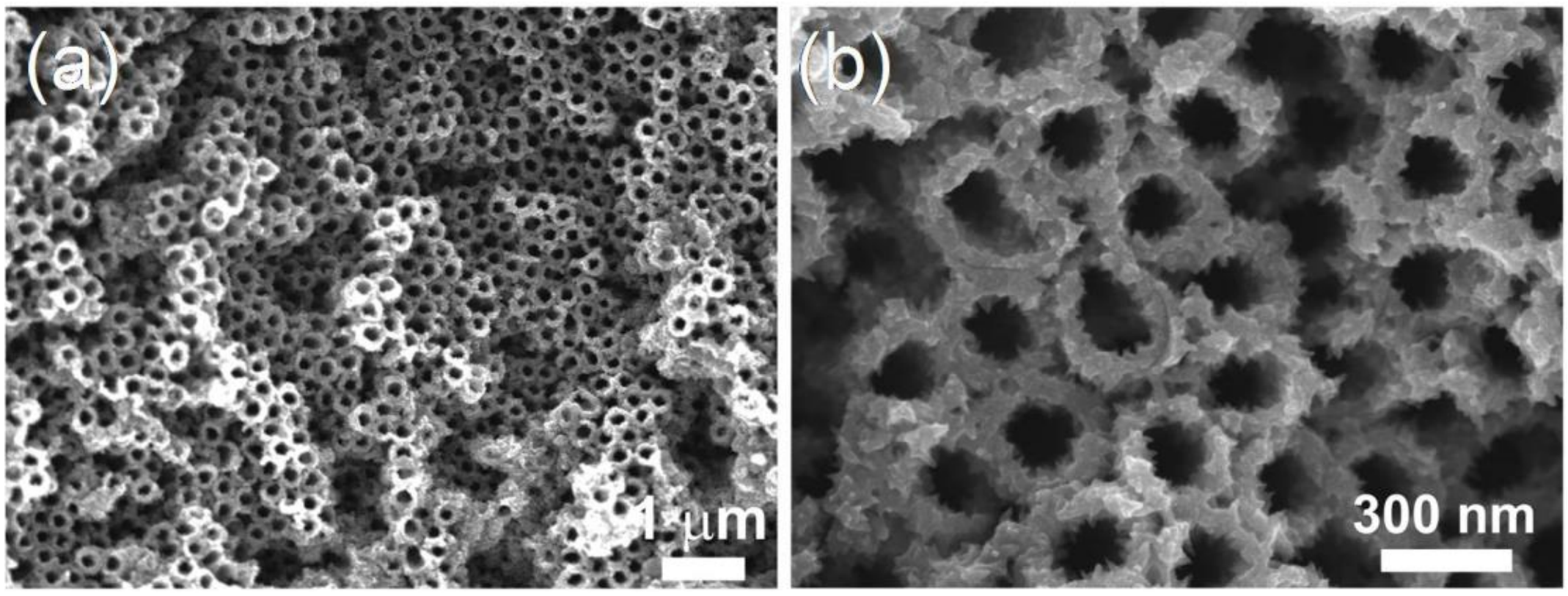

**Figure S3.** SEM images of C/TiNM: a) top view and b) magnified top view.

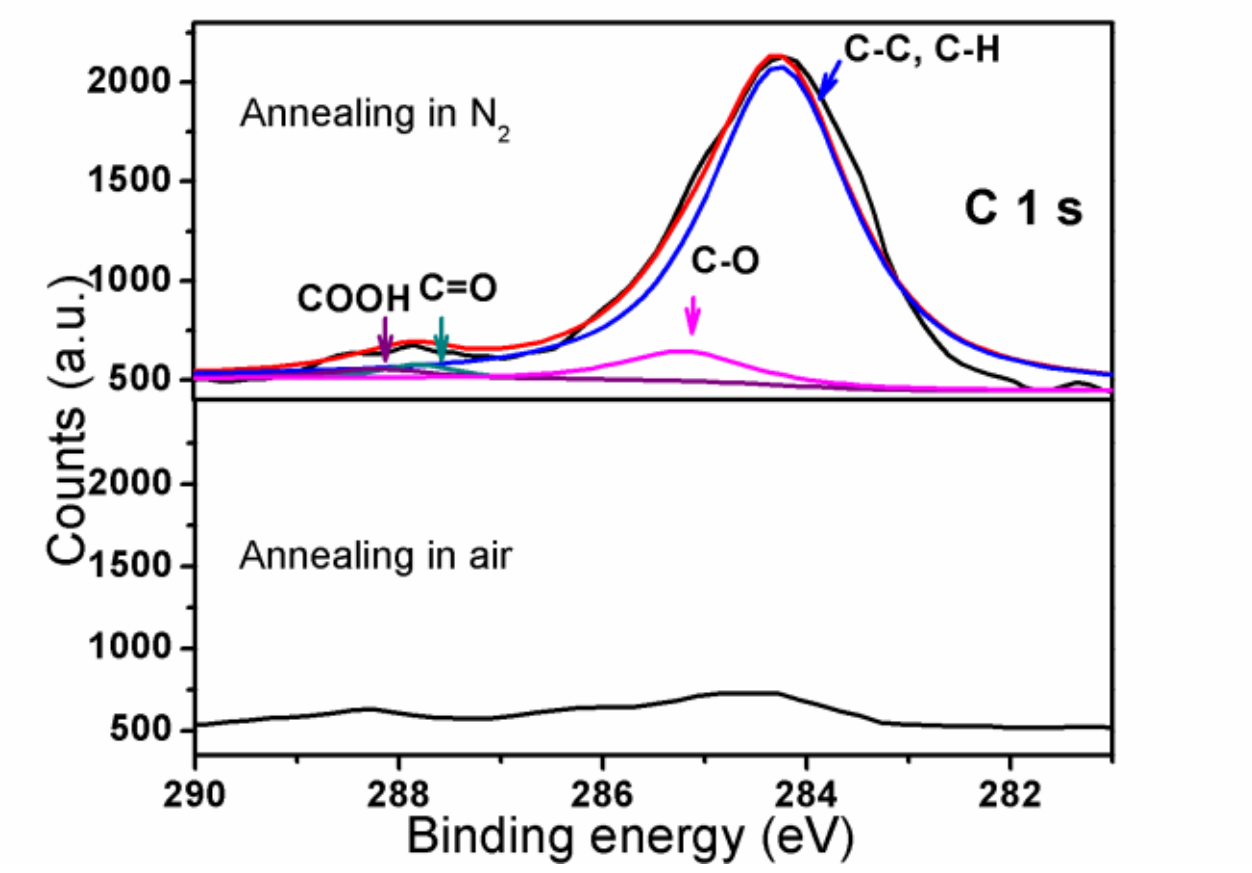

**Figure S4.** XPS high resolution C 1s peak for C/TiNM and TiNM, which prepared by annealed in $N_2$ and air, correspondingly.

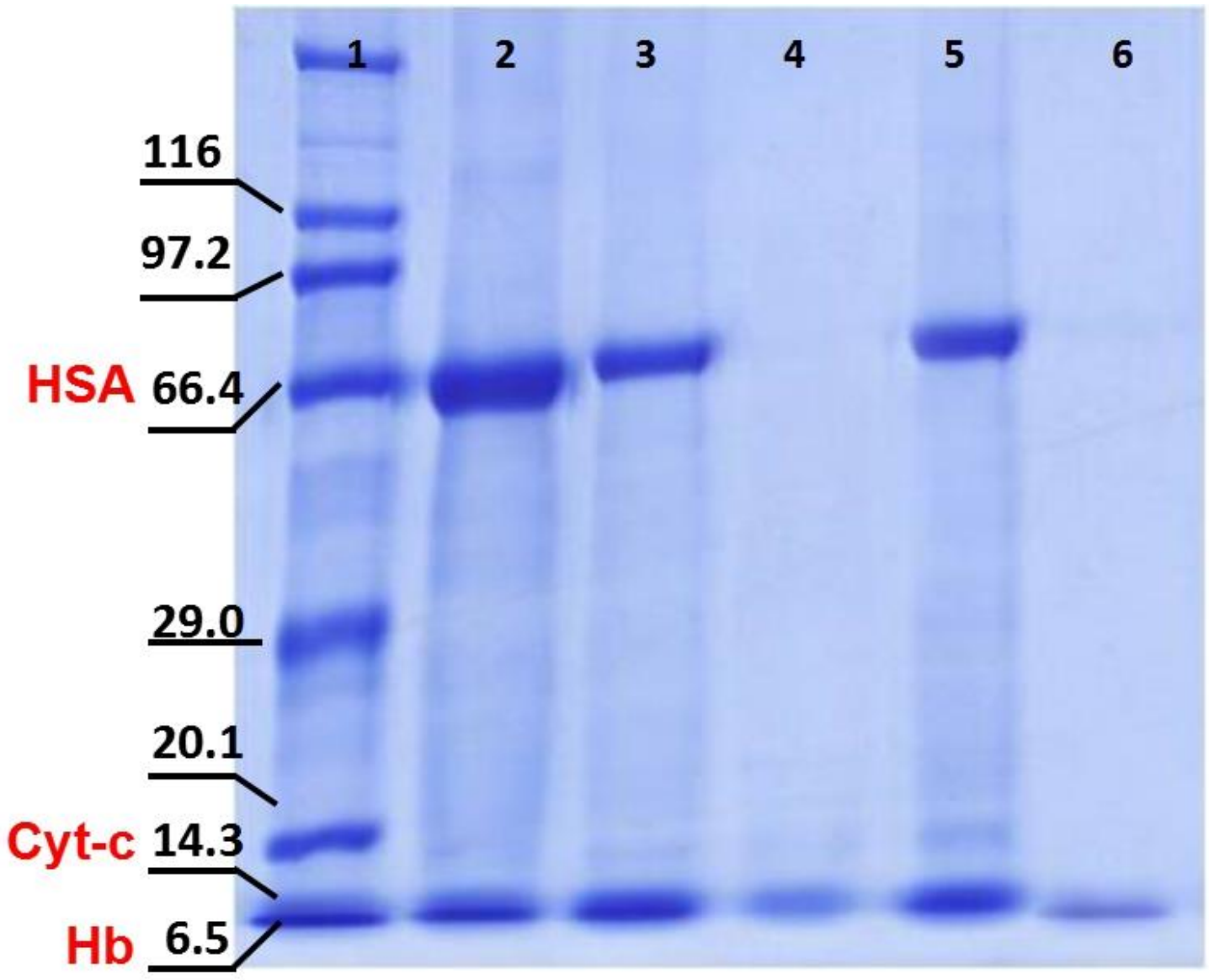

**Figure S5.** SDS-PAGE assay results: lane 1, molecular weight standards (Marker in kDa); lane 2, standard Cyt-c, Hb, and BSA mixture solutions of 400 mg $L^{-1}$, respectively; lane 3, mixture standard solutions (pH=10.1) after Cyt-c was absorbed by C/TiNM; lane 4, Recovered Cyt-c from C/TiNM (pH=3.7); lane 5, mixture standard solutions (pH=7.8) after Hb was absorbed by C/TiNM; lane 6, recovered Hb from C/TiNM (pH=3.7).

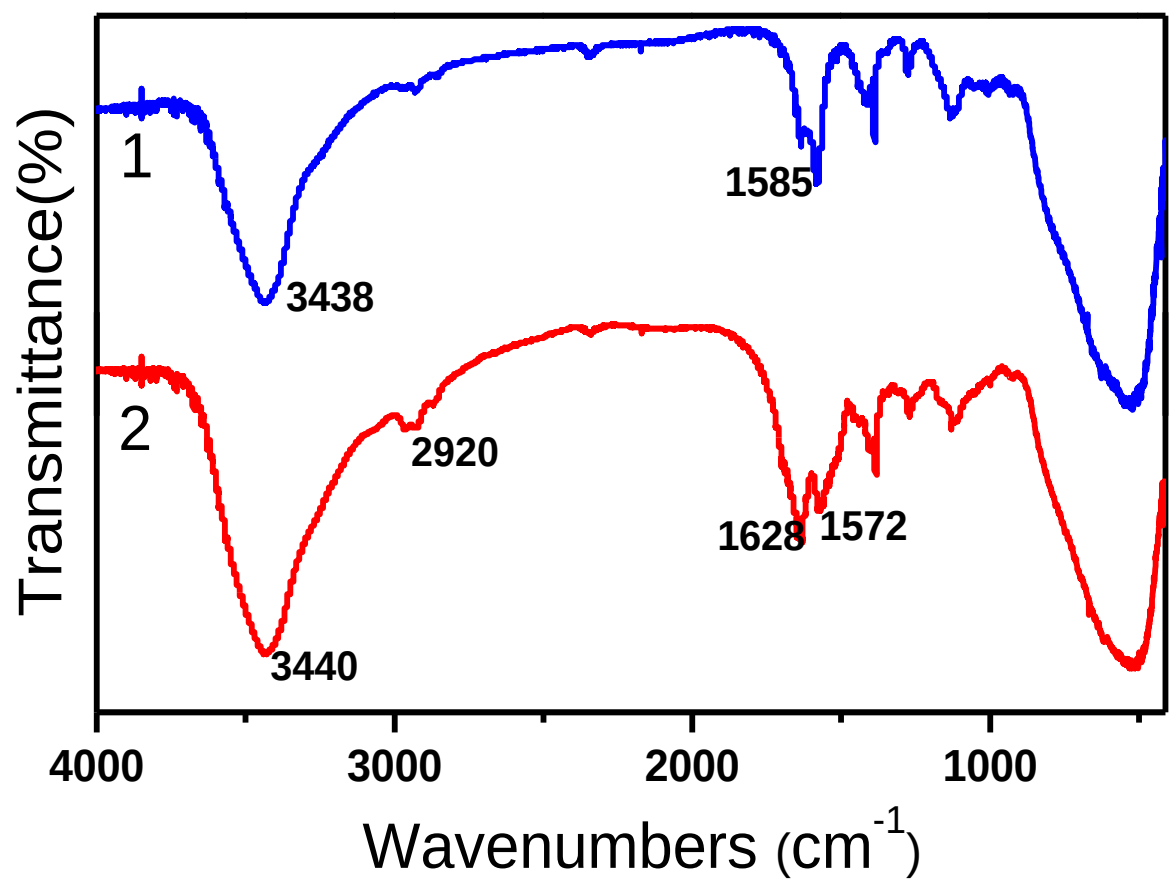

**Figure S6.** FT-IR spectra of C/TiNM before (curve 1) and after (curve 2) the adsorption of HSA.

Binding of HSA on C/TiNM can be observed in FT-IR spectra (Fig. S6) – this is most evident from two new bands that appear at around 1628 $cm^{-1}$ and 1572 $cm^{-1}$ of the C/TiNM after the adsorption of HSA molecules. These bands correspond to the C-O stretching mode of the carboxylic acid groups on side chain of peptide (amide I, 1628 $cm^{-1}$), and a combination of the N-H bending and C-N stretching of the amide plane in the backbone of protein (amide II, 1572 $cm^{-1}$), respectively. The peak at ~3440 $cm^{-1}$ can be assigned to the –OH vibrations on the surface of C/TiNM, and new band at 2920 $cm^{-1}$in curve 2 is attributed to the C–H stretching vibrations in protein.

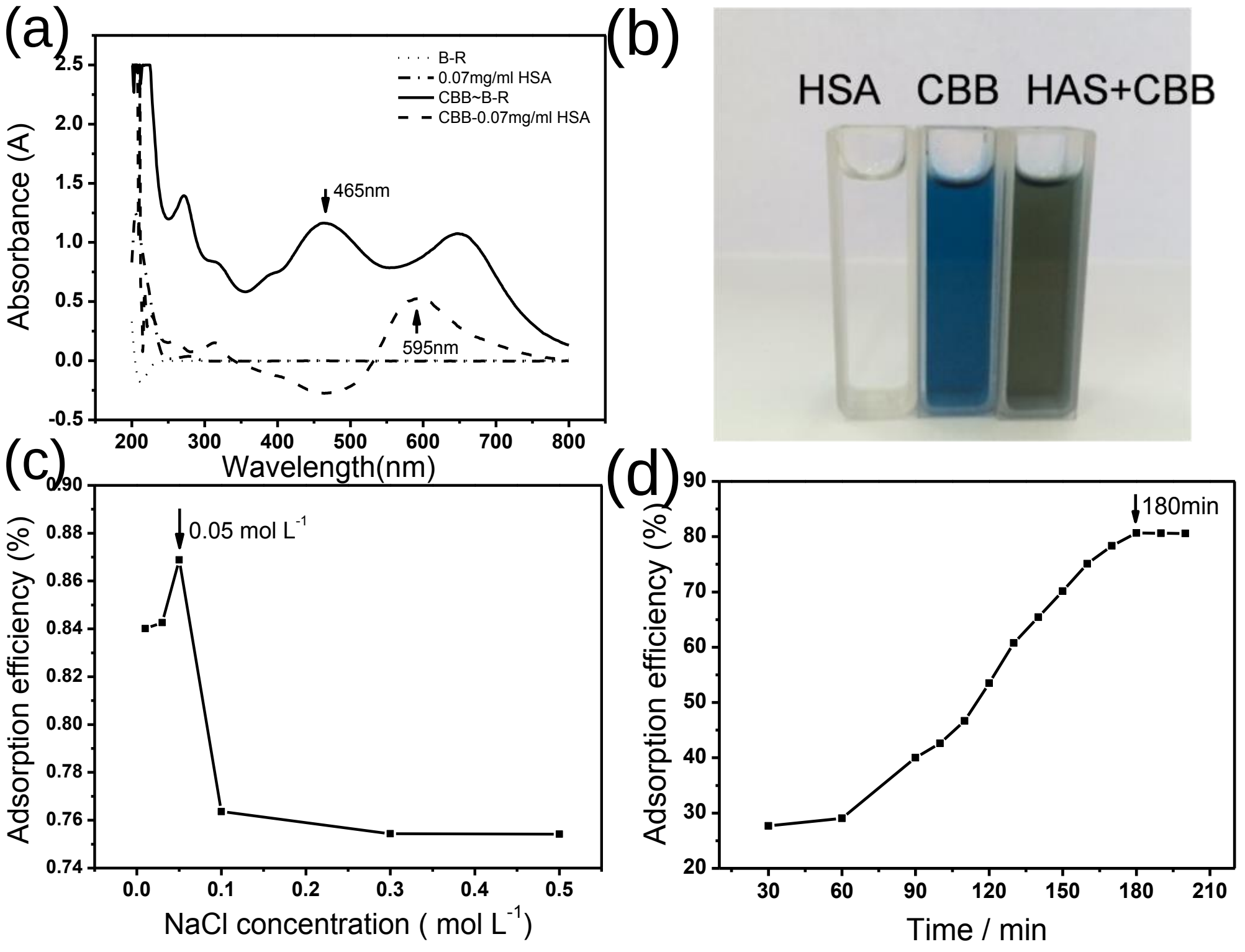

**Figure S7.** a) UV-vis spectra of HSA before and after dying by coomassie brilliant blue (CBB); b) the color of HSA, CBB, and HSA after dying by coomassie brilliant blue ; c) effect of NaCl concentration on the adsorption of HSA; and d) adsorption kinetics of C/TiNM. The HSA adsorption measurements were carried out in B-R buffer (pH=5.0) containing HSA of 140 mg $L^{-1}$, and the amount of C/TiNM: 3.0 mg.

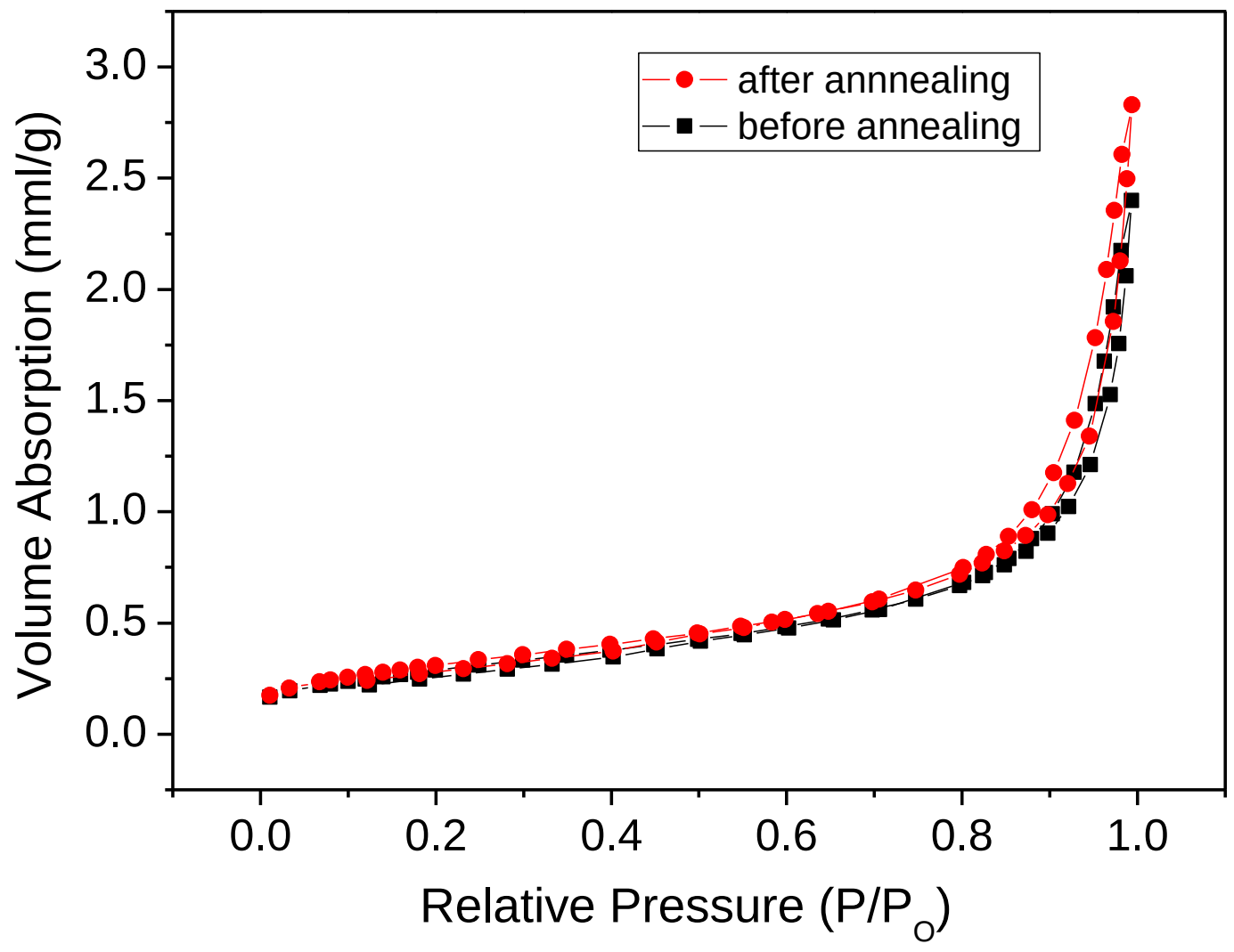

**Figure S8.** Nitrogen adsorption−desorption isotherm of TiNM before and after annealing.

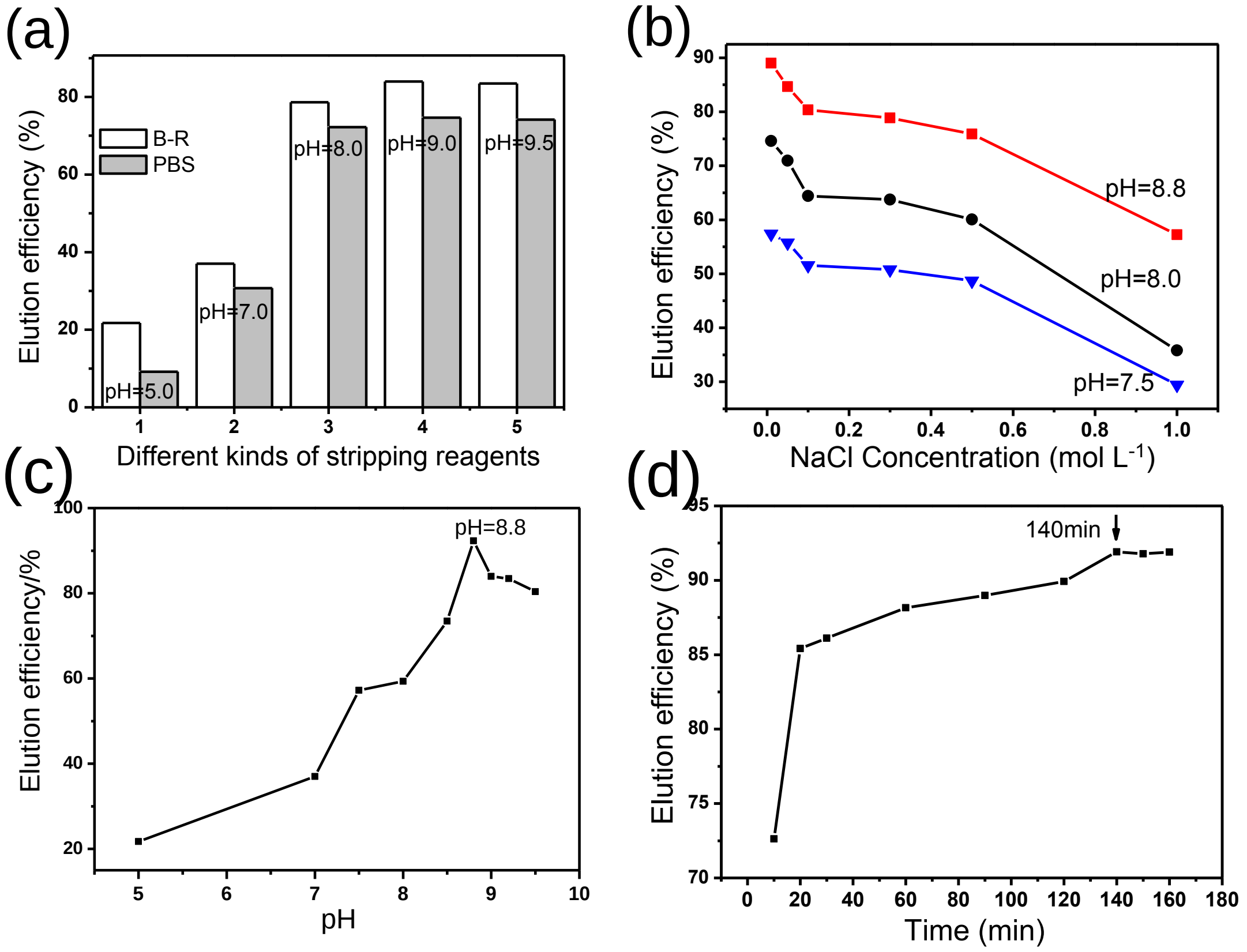

**Figure S9.** a) The comparison of elution efficiency of HSA from C/TiNM by using B-R and PBS stripping reagents; b) the effect of ion strength of B-R buffer solution on elution efficiency; c) the influence of pH value of B-R buffer solution on elution efficiency; d) effect of time on the elution efficiency.

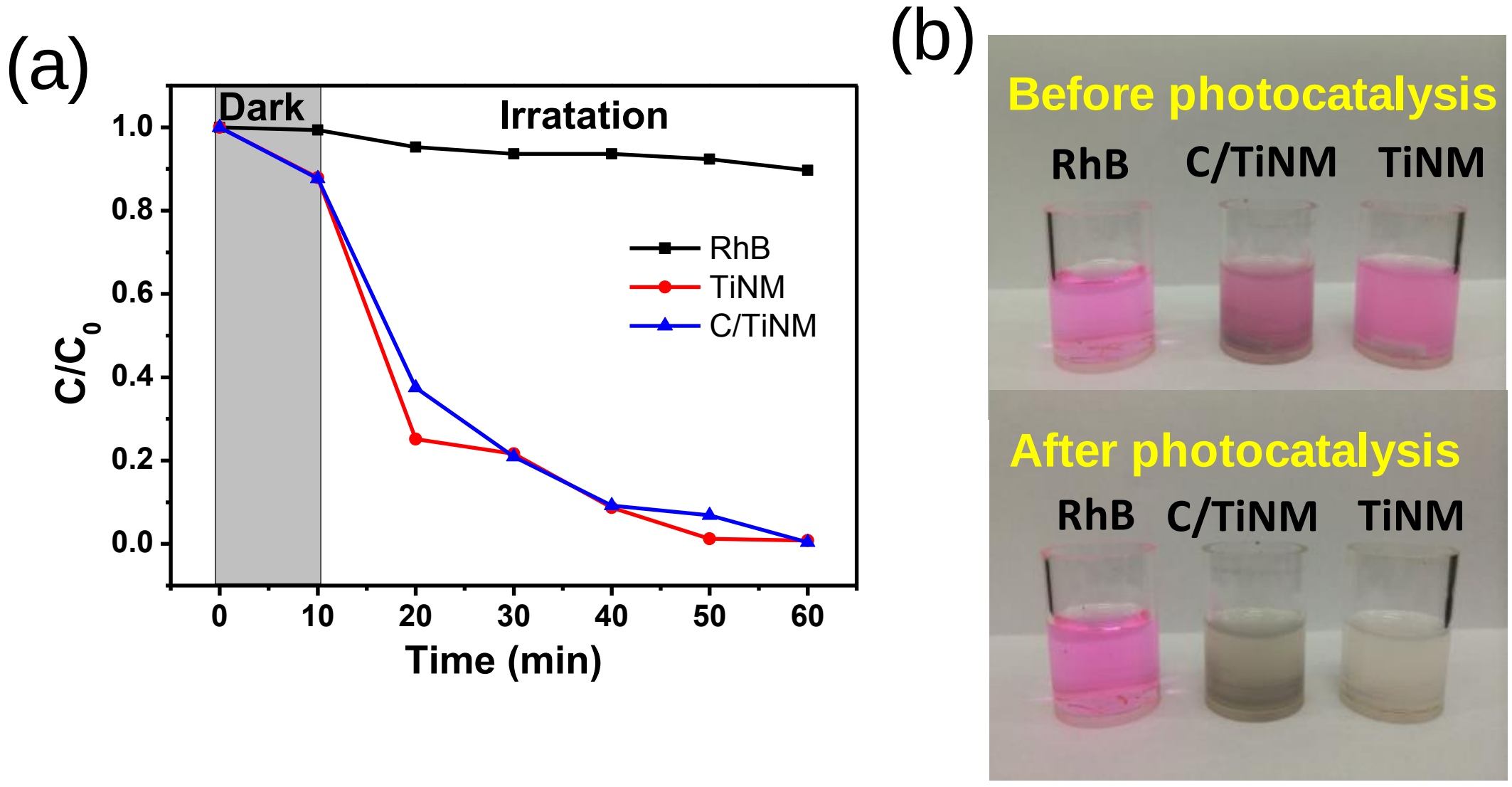

**Figure S10.** a) Photocatalytic degradation of RhB by using TiNM and C/TiNM as photocatalyst; b) digital images of RhB before and after the photocatalytical degradation under UV light irradiation without photocatalyst (left) and by using C/TiNM (middle) and TiNM (right) as photocatalyst.